\documentclass[twocolumn,prl]{revtex4}
\usepackage{float}
\usepackage{amsfonts}
\usepackage{makeidx}
\usepackage{graphicx}
\usepackage{amsmath,amssymb,graphics,epsfig,color,times,bbm}
\usepackage{CJK}

\begin{document}
\begin{CJK*}{GB}{gbsn}

\title{Continuous-variable multipartite unlockable bound entangled Gaussian states}

\author{Jing Zhang$^{\dagger}$(Õž¸)}

\affiliation{The State Key Laboratory of Quantum Optics Quantum
Optics Devices, Institute of Opto-Electronics, Shanxi University,
Taiyuan 030006, P.R. China}

\begin{abstract}
We investigate continuous-variable (CV) multipartite unlockable
bound-entangled states. Comparing with the qubit multipartite
unlockable bound-entangled states, CV multipartite unlockable
bound-entangled states present some new and different properties. CV
multipartite unlockable bound-entangled states may serve as a useful
quantum resource for new multiparty communication schemes. The
experimental protocol for generating CV unlockable bound-entangled
states is proposed with a setup that is at present accessible.
\end{abstract}

\maketitle
\end{CJK*}

Quantum entanglement is a striking property of composite quantum
systems that lies at the heart of the fundamental quantum
information protocols, which has led to ongoing efforts for its
quantitative and qualitative characterization. While entanglement of
pure bipartite states is well understood, the entanglement of mixed
and multipartite systems is still under intense research.
Entanglement is a very fragile resource, easily destroyed by the
decoherence processes to become mixed owing to unwanted coupling
with the environment. Therefore, it is important to know which mixed
states can be distilled to maximally entangled states from many
identical copies by means of local operations and classical
communication (LOCC). A surprising discovery in this area is that
there exist mixed entangled states from which no pure entanglement
can be distilled, and these states are called bound entangled states
\cite{thirteen}. This new class of states is between separable and
free-entangled states. Much effort has been devoted to the
characterization and detection of bound entanglement \cite{forteen}
and various properties and applications of bound entanglement have
been found. The distillability of multipartite entangled states,
however is much more complicated than that of bipartite entangled
states. Usually, a multipartite entangled state is bound entangled
if no pure entanglement can be distilled between any two parties by
LOCC when all the parties remain spatially separated from each
other. However, a multipartite bound entangled state may be unlocked
or activated. If all the parties are organized into several groups,
and we let each group join together and perform collective quantum
operations, then pure entanglement may be distilled between two
parties within some of the groups and this state will be called an
unlocked or activable bound entangled state. A famous class of
multipartite unlockable bound entangled states is the Smolin state
\cite{fifteen}, which is a four-qubit state and was generalized
recently into an even number of qubits \cite{sixteen,seventeen}.
These states have been applied in remote information concentration
\cite{eighteen}, quantum secret sharing \cite{nineteen},
superactivation \cite{twenty,twenty-one}. Especially, the link
between multipartite unlockable bound-entangled states and the
stabilizer formalism was found \cite{twenty-two}. The properties of
the multipartite unlockable bound-entangled states can be easily
explained from the stabilizer formalism. Recently, the four-qubit
unlockable bound-entangled state (Smolin state) was demonstrated
experimentally with polarization photons
\cite{twenty-three,twenty-three1} and ions \cite{twenty-three2}.

Most of the concepts of quantum information and computation have
been initially developed for discrete quantum variables, in
particular two-level or spin-$\frac{1}{2}$ quantum variables
(qubits). In parallel, quantum variables with a continuous spectrum,
such as the position and momentum of a particle or amplitude and
phase quadrature of an electromagnetic field, in informational or
computational processes have attracted a lot of interest and appear
to yield very promising perspectives concerning both experimental
realizations and general theoretical insights
\cite{twenty-four,twenty-five}, due to relative simplicity and high
efficiency in the generation, manipulation, and detection of CV
states. Bound entanglement of bipartite states has also been
considered for continuous variables and the nontrivial examples of
bound entangled states for CV have been constructed
\cite{twenty-six,twenty-seven,twenty-eight,twenty-eight1}. However,
The research of CV bound entanglement far lag DV. In this paper, we
first exploit the stabilizer formalism to study the CV multipartite
unlockable bound-entangled states. Comparing with the qubit
multipartite unlockable bound-entangled states, CV multipartite
unlockable bound-entangled states present some new and different
properties. We also study the four-mode multipartite unlockable
bound-entangled states in detail and present the experimental
protocol for generating CV unlockable bound entangled states.

For CV systems, the Weyl-Heisenberg group \cite{twenty-nine}, which
is the group of phase-space displacements, is a Lie group with
generators $\hat{x}= (\hat{a}+\hat{a}^{\dagger})/\sqrt{2}$
(quadrature-amplitude or position) and $\hat{p}=
-i(\hat{a}-\hat{a}^{\dagger})/\sqrt{2}$ (quadrature-phase or
momentum) satisfying the canonical commutation relation
$[\hat{x},\hat{p}] = i$ (with $\hbar = 1$). The single mode Pauli
operators (so termed in analogy with qubit systems) are defined as
$X(s) = exp[-is\hat{p}]$ and $Z(t) = exp[it\hat{x}]$ with $s,t\in
\mathbb{R}$. The Pauli operator $X(s)$ is a position-translation
operator, which acts on the computational basis of position
eigenstates $\{|q\rangle; q\in \mathbb{R}\}$ as
$X(s)|q\rangle=|q+s\rangle$, whereas $Z$ is a momentum-translation
operator, which acts on the momentum eigenstates as
$Z(t)|p\rangle=|p+t\rangle$. These operators are non-commutative and
obey the identity $ X(s)Z(t)=e^{-ist}Z(t)X(s)$. The Pauli operators
for one mode can be used to construct a set of Pauli operators
$\{X_{i}(s_{i}),Z_{i}(t_{i}); i=1,...,n\}$ for n-mode systems. The
n-mode Pauli group is expressed
\begin{equation}
G_{n} = \{ X_{1}(s_{1})Z_{1}(t_1) \otimes \dots \otimes
X_{n}(s_{n})Z_{n}(t_{n}) : s_{i}, t_{i} \in \mathbb{R}
 \}.
\end{equation}
The elements of this group can expressed in terms of a linear
combination of the canonical operators $\hat{x}_{i}$ and
$\hat{p}_{i}$ indexed by a vector $\mathbf{v} =
(s_{1},\dots,s_{n},t_{1},\dots,t_{n}) \in \mathbb{R}^{2n}$
\begin{equation}
U(\mathbf{v}) = \exp( i \sum_{i=1}^{n} (-s_{i} \hat{p}_{i} + t_{i}
\hat{x}_{i})).
\end{equation}
The commutative relationship between any two element operators in
n-mode Pauli group is expressed as
\begin{equation}
U(\mathbf{v}) U(\mathbf{v}') = e^{i \omega(\mathbf{v},\mathbf{v}')}
U(\mathbf{v}') U(\mathbf{v}),\label{commu}
\end{equation}
where $\omega(\mathbf{v},\mathbf{v}') = \sum_{i=1}^{n} (s'_{i} t_{i}
- s_{i} t'_{i})$.

Suppose we choose commuting operators $U_{1}(\mathbf{v}_{1})$,
$U_{2}(\mathbf{v}_{2})$,..., $U_{k}(\mathbf{v}_{k})$ from $G_{n}$,
and thus the $k$ independent vectors $\mathbf{v}_{1}$,...,
$\mathbf{v}_{k}$ must satisfy $\omega(\mathbf{v}_{i},\mathbf{v}_{j})
=0$ for all $i$, $j$ (see Eq. (\ref{commu})). Then we have an
Abelian subgroup
\begin{equation}
S = \{ U(\mathbf{u}) : \mathbf{u} = \sum_{i=1}^k a_{i}
\mathbf{v}_{i},\ a_{i} \in \mathbb{R} \}
\end{equation}
in which any two operators $U(\mathbf{u})$ and $U(\mathbf{u}')$
commute. The Abelian subgroup may be expressed by $S=\langle
U_{1}(\mathbf{v}_{1}), U_{2}(\mathbf{v}_{2}),...,
U_{k}(\mathbf{v}_{k}) \rangle$, which denote the Abelian subgroup
generated by them. A state $|\psi\rangle$ is said to be stabilized
by $S$, or $S$ is the stabilizer of $|\psi\rangle$, if
$U_{i}(\mathbf{v}_{i})|\psi\rangle=|\psi\rangle$, here
$i=1,2,...,k$. The stabilizer formalism for CV systems
\cite{thirty,thirty-one,thirty-two,thirty-three} has been used to
study the CV graph state \cite{thirty-four,thirty-five}. The Abelian
subgroup $S$ can be conveniently defined by its Lie algebra,
$S'=\langle H_{1}, H_{2},...,H_{k}\rangle$. The operators
$H_{i}=\mathbf{v}_{i} \mathbf{R}^{T}$ is the linear combination of
the canonical operators
$\mathbf{R}=(\hat{x}_{1},...,\hat{x}_{n},\hat{p}_{1},...,\hat{p}_{n})$.
Any two operators $H_{i}$ and $H_{j}$ commute. $S'$ is referred as
the nullifier of $|\psi\rangle$ since $H_{i}|\psi\rangle=0$,
$i=1,2,...,k$. Every nullifier is Hermitian and so observable. Thus,
the state $|\psi\rangle$ can be expressed in the simple nullifier
representation.

All the states stabilized by $S$ constitute a subspace denoted by
$V_{S}$. There is a unique pure state for $n$-mode system stabilized
by $S$ if $S$ has $n$ independent stabilizer generators (thus
$S=\langle U_{1}(\mathbf{v}_{1}), U_{2}(\mathbf{v}_{2}),...,
U_{n}(\mathbf{v}_{n}) \rangle$ are called a complete set of
stabilizer generators). When $S$ has $k$ independent stabilizer
generators which are less than the total mode number of $n$-mode
system, the states in the subspace $V_{S}$ will be more than one.
Therefore, the maximally mixed state over $V_{S}$ is expressed by
$\rho_{S}=P_{S}/tr(P_{S})$, where
\begin{equation}
P_{S} = \int
d\eta_{1}...d\eta_{k}U_{1}(\eta_{1}\mathbf{v}_{1})...U_{k}(\eta_{k}\mathbf{v}_{k})
\end{equation}
is the projection operator onto $V_{S}$. Note that the stabilized
subspace $V_{S}$ is the subspace spanned by the simultaneous
eigenstates of the stabilizer generator $\{U_{1}(\mathbf{v}_{1}),
U_{2}(\mathbf{v}_{2}),..., U_{k}(\mathbf{v}_{k})\}$ with the
eigenvalues $\{1,1,...,1\}$ (corresponding to simultaneous
eigenstates of the nullifier $\{ H_{1}, H_{2},...,H_{k}\}$ with the
eigenvalues $\{0,0,...,0\}$). In general, any orthogonal subspaces
$V_{S}^{\{\lambda_{1},...,\lambda_{k}\}}$ may be expressed by the
simultaneous eigenstates of $\{U_{1}(\mathbf{v}_{1}),
U_{2}(\mathbf{v}_{2}),..., U_{k}(\mathbf{v}_{k})\}$ with the
eigenvalues $\{e^{i\lambda_{1}} ,...,e^{i\lambda_{k}}\}$
(corresponding to simultaneous eigenstates of the nullifier $\{
H_{1}, H_{2},...,H_{k}\}$ with the eigenvalues
$\{\lambda_{1},...,\lambda_{k}\}$). The corresponding maximally
mixed state over $V_{S}^{\{\lambda_{1},...,\lambda_{k}\}}$ is
$\rho_{S}^{\{\lambda_{1},...,\lambda_{k}\}}=
P_{S}^{\{\lambda_{1},...,\lambda_{k}\}}/tr(P_{S}^{\{\lambda_{1},...,\lambda_{k}\}})$,
where
\begin{equation}
P_{S}^{\{\lambda_{1},...,\lambda_{k}\}}= \int d\eta_{1}...d\eta_{k}
e^{i\lambda_{1}\eta_{1}} U_{1}(\eta_{1}\mathbf{v}_{1})...e^{
i\lambda_{k}\eta_{k}}U_{k}(\eta_{k}\mathbf{v}_{k}) \label{dens-any}
\end{equation}
is the projection operator onto
$V_{S}^{\{\lambda_{1},...,\lambda_{k}\}}$. All these subspaces have
the same dimensions and form an orthogonal whole space.

A partition of n-mode system $\{M_{1},M_{2},...,M_{n}\}$ is defined
as a set of its proper subsets $\{V_{1},V_{2},...,V_{m}\}$, in which
$V_{i} \cap V_{j}=\varnothing (i \neq j)$,
$\cup_{i=1}^{m}V_{i}=\{M_{1},M_{2},...,M_{n}\}$, and $|V_{i}|$
denotes the number of modes in $V_{i}$. The $k$ independent
stabilizer generators can be split into local stabilizer generators
with respect to the partition $\{V_{1},V_{2},...,V_{m}\}$
$\{\{U^{(V_{1})}_{1}(\mathbf{v}_{1}),...U^{(V_{1})}_{k}(\mathbf{v}_{k})\},
\{U^{(V_{2})}_{1}(\mathbf{v}_{1}),...U^{(V_{2})}_{k}(\mathbf{v}_{k})\},...,\\
\{U^{(V_{m})}_{1}(\mathbf{v}_{1}),...U^{(V_{m})}_{k}(\mathbf{v}_{k})\}\}$
\begin{equation}
U^{(V_{\alpha})}_{\beta}(\mathbf{v}_{\beta}) = \exp( i \sum_{j\in
V_{\alpha}} (s_{j} \hat{p}_{j} + t_{j} \hat{x}_{j})).
\end{equation}
If all local stabilizer generators commute each other, the maximally
mixed state $\rho_{S}$ for n-mode system stabilized by
$\{U_{1}(\mathbf{v}_{1}), U_{2}(\mathbf{v}_{2}),...,
U_{k}(\mathbf{v}_{k})\}$ is said to be separable with respect to the
partition $\{V_{1},V_{2},...,V_{m}\}$ \cite{twenty-two}, which may
be rewrote with the product form
\begin{eqnarray}
\rho_{S} &=&
\int_{(\lambda^{U^{(V_{1})}_{1}}+...+\lambda^{U^{(V_{m})}_{1}})=0}
d\lambda^{U^{(V_{1})}_{1}}...d\lambda^{U^{(V_{m})}_{1}} \notag \\
&&
\int_{(\lambda^{U^{(V_{1})}_{2}}+...+\lambda^{U^{(V_{m})}_{2}})=0}
d\lambda^{U^{(V_{1})}_{2}}...d\lambda^{U^{(V_{m})}_{2}}...\notag \\
&&
\int_{(\lambda^{U^{(V_{1})}_{k}}+...+\lambda^{U^{(V_{m})}_{k}})=0}
d\lambda^{U^{(V_{1})}_{k}}...d\lambda^{U^{(V_{m})}_{k}}\notag \\
&&
\rho_{S^{(V_{1})}}^{\{\lambda^{U^{(V_{1})}_{1}},...,\lambda^{U^{(V_{1})}_{k}}\}}
\otimes
\rho_{S^{(V_{2})}}^{\{\lambda^{U^{(V_{2})}_{1}},...,\lambda^{U^{(V_{2})}_{k}}\}}... \notag \\
&& \otimes
\rho_{S^{(V_{m})}}^{\{\lambda^{U^{(V_{m})}_{1}},...,\lambda^{U^{(V_{m})}_{k}}\}},
\end{eqnarray}
where
$\rho_{S^{(V_{j})}}^{\{\lambda^{U^{(V_{j})}_{1}},...,\lambda^{U^{(V_{j})}_{k}}\}}$
is given in Eq. \ref{dens-any}. Moreover, if $\rho_{S}$ is separable
with respect to the partition $\{V_{1},V_{2},...,V_{m}\}$ and the
local stabilizer generators $S^{(V_{j})}=\langle
U^{(V_{j})}_{1}(\mathbf{v}_{1}),...U^{(V_{j})}_{k}(\mathbf{v}_{k})\rangle$
in one of subsets $V_{j}$ contain the number of the independent
stabilizer generators equal to the number of modes in $V_{j}$
($S^{(V_{j})}$ is a complete set of stabilizer generators on
$V_{j}$), pure entanglement among the modes inside $V_{j}$ can be
distilled \cite{twenty-two} by letting the modes in all other
subsets $V_{1},V_{2},...,V_{i\neq j},...,V_{m}$ join together and
performing joint measurements. Thus the maximally mixed state
$\rho_{S}$ for n-mode system stabilized subspace $V_{S}$ is called
an unlockable bound entangled state.

Now we consider a four-mode system with two independent stabilizers
\begin{eqnarray}
H_{1}&=& \hat{x}_{1}+\hat{x}_{2}+\hat{x}_{3}+\hat{x}_{4}\rightarrow 0, \notag \\
(U_{1}&=&Z_{1}(1)Z_{2}(1)Z_{3}(1)Z_{4}(1)\rightarrow 1), \notag \\
H_{2}&=&\hat{p}_{1}-\hat{p}_{2}+\hat{p}_{3}-\hat{p}_{4}\rightarrow 0,\notag \\
(U_{2}&=&X_{1}(1)X_{2}(-1)X_{3}(1)X_{4}(-1)\rightarrow
1),\label{four}
\end{eqnarray}
which is analogous to the four-qubit unlockable bound-entangled
state, also called Smolin state \cite{fifteen}. However, CV
four-mode unlockable bound-entangled state has some distinct
properties comparing with the counterpart of qubit. Considering the
$2:2$ partition $\{\{M_{1},M_{2}\},\{M_{3},M_{4}\}$, we have local
stabilizer generators
$\{\{U_{1}^{(\{1,2\})}=Z_{1}(1)Z_{2}(1),U_{2}^{(\{1,2\})}=X_{1}(1)X_{2}(-1)\},
\{U_{1}^{(\{3,4\})}=Z_{3}(1)Z_{4}(1),U_{2}^{(\{3,4\})}=X_{3}(1)X_{4}(-1)\}$,
which commute each other. Therefore the maximally mixed state
$\rho_{S}^{(4)}$ stabilized by $U_{1}$ and $U_{2}$ may be expressed
by the product form with respect to the partition
$\{\{M_{1},M_{2}\},\{M_{3},M_{4}\}$
\begin{eqnarray}
\rho_{S}^{(4)} &=&\frac{1}{tr(P_{S})}\int
d\eta_{1}d\eta_{2}Z_{1}(\eta_{1})Z_{2}(\eta_{1})Z_{3}(\eta_{1})Z_{4}(\eta_{1})\notag \\
&& X_{1}(\eta_{2})X_{2}(-\eta_{2})X_{3}(\eta_{2})X_{4}(-\eta_{2}) \notag \\
&=&\int d\lambda_{1} d\lambda_{2}
\rho_{S^{(\{M_{1},M_{2}\})}}^{\{\lambda_{1},\lambda_{2}\}}\otimes
\rho_{S^{(\{M_{3},M_{4}\})}}^{\{-\lambda_{1},-\lambda_{2}\}},\label{four1}
\end{eqnarray}
where $\rho_{S^{(\{M_{1},M_{2}\})}}^{\{\lambda_{1},\lambda_{2}\}}=
\int d\eta_{1} d\eta_{2} e^{i\lambda_{1}\eta_{1}}
Z_{1}(\eta_{1})Z_{2}(\eta_{1}) e^{i\lambda_{2}\eta_{2}} \\
X_{1}(\eta_{2})X_{2}(-\eta_{2})$ and
$\rho_{S^{(\{M_{3},M_{4}\})}}^{\{-\lambda_{1},-\lambda_{2}\}}$ is
similar as
$\rho_{S^{(\{M_{1},M_{2}\})}}^{\{\lambda_{1},\lambda_{2}\}}$. The
$\rho_{S}^{(4)}$ is separable with respect to the $2:2$ partition
$\{\{M_{1},M_{2}\},\{M_{3},M_{4}\}\}$. Furthermore, we consider the
$2:2$ partition $\{\{M_{1},M_{4}\},\{M_{2},M_{3}\}\}$, whose
properties are the same as the partition
$\{\{M_{1},M_{2}\},\{M_{3},M_{4}\}\}$. However, the partition
$\{\{M_{1},M_{3}\},\{M_{2},M_{4}\}\}$ is quite different, since its
local stabilizer generators don't commute in the same subset. Thus,
the $\rho_{S}^{(4)}$ is inseparable with respect to the partition
$\{\{M_{1},M_{3}\},\{M_{2},M_{4}\}\}$. Comparing the CV four-mode
unlockable bound-entangled state, the four-qubit unlockable
bound-entangled state is invariant under arbitrary permutation of
the four qubits and is separable with respect to any $2:2$
partition.

$ Nondistillabitily:$ When the four parties sharing four modes
respectively are located in separated stations (thus they can not
perform joint quantum operation), they can not distill any pure
entanglement by LOCC. This comes from the fact the state is
separable with respect to the partitions
$\{\{M_{1},M_{2}\},\{M_{3},M_{4}\}\}$ and
$\{\{M_{1},M_{4}\},\{M_{2},M_{3}\}\}$. In detail, since the state is
separable across $\{\{M_{1},M_{2}\},\{M_{3},M_{4}\}\}$, local
measurements and classical communication for $M_{1}$ and $M_{3}$
($M_{1}$ and $M_{4}$; $M_{2}$ and $M_{3}$; $M_{2}$ and $M_{4}$) can
not establish any entanglement between $M_{2}$ and $M_{4}$ ($M_{2}$
and $M_{3}$; $M_{1}$ and $M_{4}$; $M_{1}$ and $M_{3}$) respectively
(since the amount of entanglement can not be increased by local
operations and classical communication
\cite{thirty-six,thirty-seven}). Considering the state is separable
across $\{\{M_{1},M_{4}\},\{M_{2},M_{3}\}\}$, local measurements and
classical communication for $M_{1}$ and $M_{2}$ ($M_{1}$ and
$M_{3}$; $M_{4}$ and $M_{2}$; $M_{4}$ and $M_{3}$) can not establish
any entanglement between $M_{4}$ and $M_{3}$ ($M_{4}$ and $M_{2}$;
$M_{1}$ and $M_{3}$; $M_{1}$ and $M_{2}$) respectively. Thus, by
definition this state is called a multi-partite bound-entangled
state.

$ Unlockability:$ Though this state is nondistillable under LOCC
when all the modes remain spatially separated from each other, its
entanglement can be unlocked. For example, considering the state is
separable across $\{\{M_{1},M_{2}\},\{M_{3},M_{4}\}\}$, performing
the joint Bell-basis measurement on $M_{1}$ and $M_{4}$ ($M_{2}$ and
$M_{3}$) can establish pure entanglement between $M_{2}$ and $M_{3}$
($M_{1}$ and $M_{4}$) respectively (easily see Eq. (9)). However,
performing the joint Bell-basis measurement on $M_{1}$ and $M_{3}$
($M_{2}$ and $M_{4}$) can not establish any entanglement between
$M_{2}$ and $M_{4}$ ($M_{1}$ and $M_{3}$) respectively since the
local stabilizer generators of $\{\{M_{1},M_{3}\},\{M_{2},M_{4}\}\}$
don't commute in the same subset. Considering that the state is
separable across $\{\{M_{1},M_{4}\},\{M_{2},M_{3}\}\}$, the joint
Bell-basis measurement performing on $M_{1}$ and $M_{2}$ ($M_{4}$
and $M_{3}$) can establish pure entanglement between $M_{4}$ and
$M_{3}$ ($M_{1}$ and $M_{2}$) respectively. However, the joint
Bell-basis measurement performing on $M_{1}$ and $M_{3}$ ($M_{4}$
and $M_{2}$) can not establish any entanglement between $M_{4}$ and
$M_{2}$ ($M_{1}$ and $M_{3}$) respectively.

The EPR entangled state can also be distilled by employing the
tensor product of two identical CV four-mode unlockable
bound-entangled states, which is analogous to the distillation
process for the four-qubit unlockable bound-entangled state, also
called superactivation of bound entanglement \cite{twenty}. Two
identical CV four-mode unlockable bound-entangled states
$\rho_{1,2,3,4}^{(S)}$ and $\rho_{1',2',3',4'}^{(S)}$ are assigned
to five remote parties A, B, C, D, and E as shown in Fig. 1. Thus
the parties A, B, C, and D each have one mode 1, 2, 3, and 4
respectively of state $\rho_{1,2,3,4}^{(S)}$ and similarly parties
A, B, C, E each have one mode 2', 3', 4', and 1' of state
$\rho_{1',2',3',4'}^{(S)}$. The parties A, B, and C perform joint
Bell-basis measurement respectively, then send their measured
results to D. The party D translates the measurement results into
the mode 4, which is expressed by
\begin{eqnarray}
\hat{x}_{4}'&=&\hat{x}_{4}+(\hat{x}_{1}+\hat{x}_{2'})+(\hat{x}_{2}+\hat{x}_{3'})+(\hat{x}_{3}+\hat{x}_{4'})\notag \\
&=&\hat{x}_{2'}+\hat{x}_{3'}+\hat{x}_{4'},\notag \\
\hat{p}_{4}'&=&\hat{p}_{4}-(\hat{p}_{1}-\hat{p}_{2'})+(\hat{p}_{2}-\hat{p}_{3'})-(\hat{p}_{3}-\hat{p}_{4'})\notag \\
&=&\hat{p}_{2'}-\hat{p}_{3'}+\hat{p}_{4'}.
\end{eqnarray}
Thus an EPR pair between D and E is distilled with
$\hat{x}_{1'}+\hat{x}_{4}'\rightarrow 0$ and
$\hat{p}_{1'}-\hat{p}_{4}'\rightarrow 0$.

%
\begin{figure}
\centerline{
\includegraphics[width=2in]{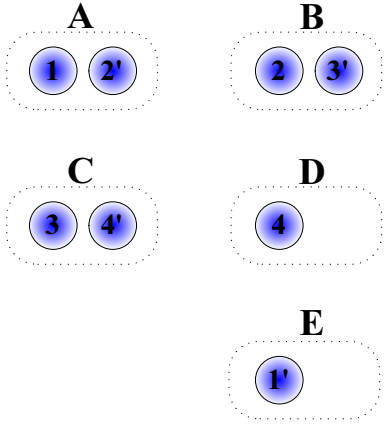}
} \vspace{0.1in}
\caption{(Color online). Distill the state
$\rho_{1,2,3,4}^{(S)}\bigotimes\rho_{1',2',3',4'}^{(S)}$ into an EPR
pair between D and E by local measurements. \label{Fig1} }
\end{figure}

The CV four-mode unlockable bound-entangled state may be generalized
into $2n$ modes, whose nullifiers (stabilizer generators) are
expressed by
\begin{eqnarray}
H_{1}^{(2n)}&=& \hat{x}_{1}+\hat{x}_{2}+\hat{x}_{3}+\hat{x}_{4}+...+\hat{x}_{2n-1}+\hat{x}_{2n}, \notag \\
H_{2}^{(2n)}&=&\hat{p}_{1}-\hat{p}_{2}+\hat{p}_{3}-\hat{p}_{4}+...+\hat{p}_{2n-1}-\hat{p}_{2n}.
\end{eqnarray}
It can be easily seen  hat the maximally mixed state
$\rho_{S}^{(2n)}$ stabilized by $U_{1}^{2n}$ and $U_{2}^{2n}$ is
separable not for any $2:2:...:2$ partition. Applying this method,
many more CV unlockable bound entangled states can be found and
defined.

%
\begin{figure}
\centerline{
\includegraphics[width=3in]{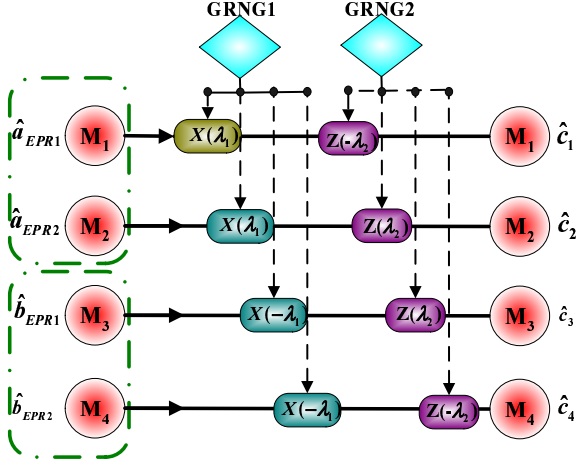}
} \vspace{0.1in}
\caption{(Color online). The generation of four-mode unlockable
bound-entangled state. GRNG: Gaussian random number generator. The
$X$ and $Z$ are the position and momentum -translation Pauli
operators respectively. \label{Fig2} }
\end{figure}

The above analyses of the CV multipartite unlockable bound-entangled
states based on the stabilizer formula, require infinite energy and
stand as an idealized limit. Now we investigate the four-mode
unlockable bound-entangled state with finite squeezing and present
the protocol to generate it experimentally. As shown in Fig. 2, two
pairs $(\hat{a}_{EPR1},\hat{a}_{EPR2})$ and
$(\hat{b}_{EPR1},\hat{b}_{EPR2})$ of EPR entangled states (or called
the two-mode squeezed state
$|\psi(r)\rangle=\sum_{n}\lambda^{n}\sqrt{1-\lambda^{2}}|n,n\rangle$
with $\lambda=\tanh r$, where $r$ is the squeezing factor) are
distributed into four stations $
M_{1}(\hat{a}_{EPR1}),M_{2}(\hat{a}_{EPR2}),M_{3}(\hat{b}_{EPR1}),M_{4}(\hat{b}_{EPR2})$
respectively. The EPR entangled state has a very strong correlation
property,
name that both their sum-amplitude quadrature variance $\langle \delta^2 (\hat{x}%
_{a(b)_{EPR1}}+\hat{x}_{a(b)_{EPR2}})\rangle =e^{-2r}$, and their
difference-phase quadrature variance $\langle \delta^2 (\hat{p}_{a(b)_{EPR1}}-%
\hat{p}_{a(b)_{EPR2}})\rangle =e^{-2r}$, are less than the quantum
noise limit \cite{Simon,Duan}. The position and momentum of two
pairs $(\hat{a}_{EPR1},\hat{a}_{EPR2})$ and
$(\hat{b}_{EPR1},\hat{b}_{EPR2})$ are translated random using two
Gaussian random number generators (GRNG) as shown in Fig. 2. This
random operation applied exhibits a Gaussian distribution, hence the
standard deviation of the GRNG $\sigma_{GRNG}$ provides a complete
characterization of its strength. The resulting state is expressed
by
\begin{eqnarray}
\hat{c}_{1}&=&\hat{a}_{EPR1}+x_{GRNG1}-p_{GRNG2},\notag \\
\hat{c}_{2}&=&\hat{a}_{EPR2}+x_{GRNG1}+p_{GRNG2},\notag \\
\hat{c}_{3}&=&\hat{b}_{EPR1}-x_{GRNG1}+p_{GRNG2},\notag \\
\hat{c}_{4}&=&\hat{b}_{EPR2}-x_{GRNG1}-p_{GRNG2},\label{experi}
\end{eqnarray}
and the correlation variances of two independent stabilizers of this
state are $\langle\delta
^{2}(\hat{x}_{c_{1}}+\hat{x}_{c_{2}}+\hat{x}_{c_{3}}+\hat{x}_{c_{4}})\rangle=2e^{-2r}$
and $\langle\delta
^{2}(\hat{p}_{c_{1}}-\hat{p}_{c_{2}}+\hat{p}_{c_{3}}-\hat{p}_{c_{4}})\rangle=2e^{-2r}$.
The output state will exactly become that in Eq. \ref{four} (and be
expressed by the density operator Eq. \ref{four1}) when
$r\rightarrow\infty$ and $\sigma_{GRNG}\rightarrow\infty$. Note that
the $\lambda_{1}$ and $\lambda_{2}$ for GRNG 1 and 2 in Fig. 2
correspond to that of the density operator $\rho_{S}^{(4)}$ of Eq.
10. When $\lambda_{1}=\lambda_{2}=0$, it corresponds to two original
input pairs of EPR entangled state without performing GRNG, which
may be expressed by the density operator
$\rho_{S^{(\{M_{1},M_{2}\})}}^{\{0,0\}}=\int d\eta_{1} d\eta_{2}
Z_{1}(\eta_{1})Z_{2}(\eta_{1})X_{1}(\eta_{2})X_{2}(-\eta_{2})$ (and
$\rho_{S^{(\{M_{3},M_{4}\})}}^{\{0,0\}}$ is similar as
$\rho_{S^{(\{M_{1},M_{2}\})}}^{\{0,0\}}$ ) or
$\hat{x}_{a(b)_{EPR1}}+\hat{x}_{a(b)_{EPR2}}
=\hat{p}_{a(b)_{EPR1}}-\hat{p}_{a(b)_{EPR2}} \rightarrow0$. Here,
due to finite squeezing, the strength of the GRNG don't need to be
infinite, but will have lower limit value depending on the squeezing
factor $r$.

\begin{figure}
\centerline{
\includegraphics[width=3in]{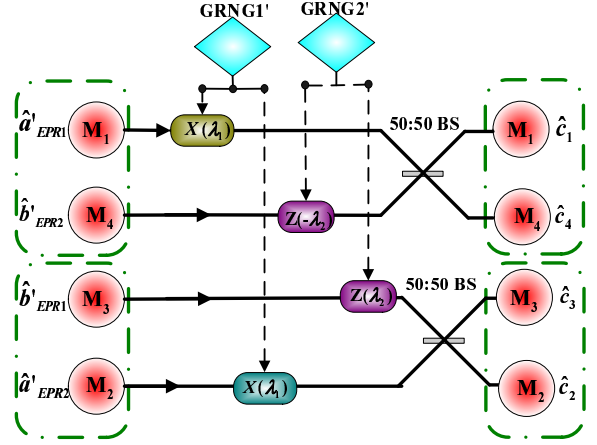}
} \vspace{0.1in}
\caption{(Color online). The four-mode unlockable bound-entangled
state in Fig.1 is generated equivalently with considering the
partition $\{\{M_{1},M_{4}\},\{M_{2},M_{3}\}$. BS: Beamsplitter.
\label{Fig3} }
\end{figure}

Giedke et. al \cite{Giedke} give a necessary and sufficient
condition for separability of Gaussian states of bipartite systems
of arbitrarily many modes. The condition provides an operational
criterion since it can be checked by simple computation with the
covariance matrix (CM). The Wigner distribution of the Gaussian
states can be constructed as $W(R)=\pi^{-N} \exp[-R^T \cdot
\Gamma^{-1} \cdot R]$, where
$R=(x_1,\,p_1,\,x_2,\,p_2,\,\ldots,\,x_N,\,p_N)^T$ is the vector of
phase-space variables. This implicitly defines the elements of the
CM $\Gamma$, which up to local displacements provides a complete
description of the Gaussian states \cite{twenty-four,twenty-five}.
Thus CM of the four-mode unlockable bound-entangled state with
finite squeezing can be derived by Eq. (13) and may be used to
achieve the condition for separability \cite{Giedke}.

\begin{figure}
\centerline{
\includegraphics[width=3in]{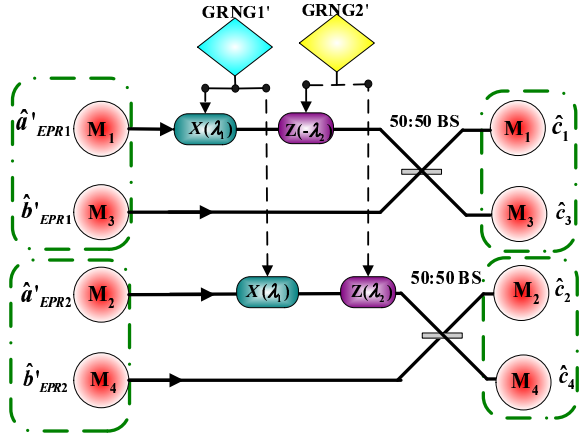}
} \vspace{0.1in}
\caption{(Color online). The four-mode unlockable bound-entangled
state in Fig. 2 is generated equivalently with considering the
partition $\{\{M_{1},M_{3}\},\{M_{2},M_{4}\}$. \label{Fig4} }
\end{figure}

The state $\hat{c}_{1},\hat{c}_{2},\hat{c}_{3},\hat{c}_{4}$ with
respect to the partition $\{\{M_{1},M_{2}\},\{M_{3},M_{4}\}$ is
always separable not depending on the strength of the GRNG as seen
in Fig. 2, since the amount of entanglement can not be increased by
local operations and classical communication
\cite{thirty-six,thirty-seven}. However, the separability of the
state $\hat{c}_{1},\hat{c}_{2},\hat{c}_{3},\hat{c}_{4}$ with respect
to the partition $\{\{M_{1},M_{4}\},\{M_{2},M_{3}\}$ depends on the
strength of the GRNG. The state
$\hat{c}_{1},\hat{c}_{2},\hat{c}_{3},\hat{c}_{4}$ can be generated
equivalently as shown in Fig. 3. Thus, we may utilize directly the
separability criterion in term of measurable squeezing variances of
two-mode states
\begin{eqnarray}
\langle\delta ^{2}(\hat{x}_{1}+\hat{x}_{2})\rangle +\langle\delta
^{2}(\hat{p}_{1}-\hat{p}_{2})\rangle\geq 2 . \label{sep}
\end{eqnarray}
This is sufficient criterion for separability for any two-mode
state, expressed in a form suitable for experimental verification
\cite{Simon,Duan}. The lowest limit value of the strength of the
GRNG1' can be obtained by
\begin{eqnarray}
\langle\delta ^{2}(\hat{x}_{a'_{EPR1}}+\hat{x}_{a'_{EPR2}})\rangle
+4\langle\delta ^{2}(\hat{x}_{GRNG1'})\rangle+ \notag \\
\langle\delta
^{2}(\hat{p}_{a'_{EPR1}}-\hat{p}_{a'_{EPR2}})\rangle\geq 2
\Rightarrow \notag \\
\langle\delta ^{2}(\hat{x}_{GRNG1'})\rangle\geq (1-e^{-2r})/2.
\label{sep}
\end{eqnarray}
The lowest limit value of the strength of the GRNG2' can be also
obtained with $\langle\delta ^{2}(\hat{p}_{GRNG2'})\rangle\geq
(1-e^{-2r})/2$. Note that whether this lowest value for GRNG is the
necessary and sufficient condition for separability of the state
$\hat{c}_{1},\hat{c}_{2},\hat{c}_{3},\hat{c}_{4}$ with respect to
the partition $\{\{M_{1},M_{4}\},\{M_{2},M_{3}\}$, still must be
further studied by applying the necessary and sufficient condition
for separability of Gaussian states of bipartite systems of
arbitrarily many modes to check the separability \cite{Giedke}.
Considering the partition $\{\{M_{1},M_{3}\},\{M_{2},M_{4}\}$, the
state $\hat{c}_{1},\hat{c}_{2},\hat{c}_{3},\hat{c}_{4}$ in Fig. 2
can be generated equivalently as shown in Fig. 4. It is easily seen
that there is an EPR entangled states without any translation
operations. Therefore the four-mode unlockable bound-entangled state
is always entangled with respect to the partition
$\{\{M_{1},M_{3}\},\{M_{2},M_{4}\}$. Moreover, the EPR entanglement
between mode 1 and 2 (or 3 and 4) can be distilled by letting mode 3
and 4 (1 and 2) come together and performing the joint Bell-basis
measurement and the resulting EPR entanglement becomes $\langle \delta^2 (\hat{x}%
_{EPR1(3)}+\hat{x}_{EPR2(4)})\rangle =2e^{-2r}$ and $\langle \delta^2 (\hat{p}_{EPR1(3)}-%
\hat{p}_{EPR2(4)})\rangle =2e^{-2r}$ (here the gain factor is 1).
However, the EPR entanglement between mode 1 and 3 (or 2 and 4) can
not be distilled by letting mode 2 and 4 (1 and 3) come together and
performing the joint Bell-basis measurement.

Here, we would like to emphasize that the definition of the
multi-partite bound-entangled state in this paper is completely
different from the bipartite bound entangled Gaussian states defined
in Ref. \cite{twenty-seven}. In Ref. \cite{twenty-seven}, they study
the bound entangled Gaussian states for bipartite system with
arbitrary number of modes in each party. The bipartite bound
entangled Gaussian states are positive partial transpose, thus are
undistillable, and they are not separable. According to the
definition of the bipartite bound entangled Gaussian states defined
in Ref. \cite{twenty-seven}, the CV four-mode unlockable
bound-entangled state in this paper has three possibilities for
bipartition: $\{\{M_{1},M_{2}\},\{M_{3},M_{4}\}\}$,
$\{\{M_{1},M_{4}\},\{M_{2},M_{3}\}\}$, and
$\{\{M_{1},M_{3}\},\{M_{2},M_{4}\}\}$. The bipartition
$\{\{M_{1},M_{2}\},\{M_{3},M_{4}\}\}$ (and
$\{\{M_{1},M_{4}\},\{M_{2},M_{3}\}\}$) of CV four-mode unlockable
bound-entangled state is separable, which does not hold for a
bipartite bound entangled Gaussian state. The bipartition
$\{\{M_{1},M_{3}\},\{M_{2},M_{4}\}\}$ of CV four-mode unlockable
bound-entangled state is inseparable and also nonpositive partial
transpose, thus it is entangled and distillable.

In conclusion, we have introduced CV multipartite unlockable
bound-entangled states. It is interesting to further investigate the
relationship between the finite squeezing and the strength of the
GRNG for more complex CV multipartite unlockable bound-entangled
states, which relates to separability problem. CV multipartite
unlockable bound-entangled states may serve as a useful quantum
resource for new multiparty communication schemes in the
continuous-variable field, such as remote information concentration
quantum secret sharing, superactivation. We believe that this work
here will contribute to deeper understanding of CV entanglement.

$\dagger$Corresponding author's email address: jzhang74@yahoo.com,
jzhang74@sxu.edu.cn

\textbf{ACKNOWLEDGMENTS}

J.Z. thanks K. Peng, C. Xie, and P. van Loock for the helpful
discussions. This research was supported in part by NSFC for
Distinguished Young Scholars (Grant No. 10725416), National Basic
Research Program of China (Grant No. 2011CB921601), NSFC Project for
Excellent Research Team (Grant No. 60821004).

\end{document}